\documentclass[11pt,a4paper]{article}
\usepackage[body={16.5cm,25cm}]{geometry}
\usepackage[dvips]{color}
\usepackage{array}
\usepackage{stackrel}
\usepackage{tensor}
\usepackage{mathtools}
\usepackage{curves}
\usepackage[titletoc]{appendix}
\usepackage{setspace}
\usepackage[mathscr]{eucal}

\usepackage[hang,small,center]{caption}
\usepackage{amsmath,amssymb}
\usepackage{amsfonts}
\usepackage{mathrsfs}
\usepackage{epic}
\usepackage{eepic}
\usepackage{graphicx}
\usepackage{cite}
\newcommand{\ds}{\displaystyle}

\renewcommand{\author}[1]{\large\rm #1\\ \bigskip}
\newcommand{\address}[1]{{\normalsize\it #1\\}\bigskip}
\renewcommand{\title}[1]{\bigskip\bigskip\Large\bf #1\bigskip\bigskip\\}

\newcommand{\Bigpsi}[3]{\phantom{\Psi}_2 \kern -.05em
\Psi_2\left(\genfrac{}{}{0pt}{}{#1}{#2}\biggl|#3\right)}

\def\bea{\begin{eqnarray}}
\def\eea{\end{eqnarray}}
\newcommand{\beq}{\begin{equation}}
\newcommand{\eeq}{\end{equation}}

\newcommand{\ml}[1]{\mathcal{#1}}

\newcommand{\lam}{{\lambda}}

\newcommand{\phifs}{\tensor*[_{4}]{{\phi}}{_3}}
\newcommand{\phit}{\tensor*[_{3}]{\overline{\phi}}{_2}}
\newcommand{\phis}{\tensor*[_{2}]{\overline{\phi}}{_1}}

\newcommand{\phir}{\tensor*[_{r+1}]{\overline{\phi}}{_r}}
\newcommand{\phirs}{\tensor*[_{r+1}]{{\phi}}{_r}}

\newcounter{app}
\newcounter{sapp}[app]

\def\nsection#1{\setcounter{equation}{0}\section{#1}}

\begin{document}

\vglue 2 cm
\begin{center}
\title{$Q$-operators in the six-vertex model}
\author{Vladimir
  V.~Mangazeev$^{1,2}$ }
\address{$^1$Department of Theoretical Physics,
         Research School of Physics and Engineering,\\
    Australian National University, Canberra, ACT 0200, Australia.\\\ \\
$^2$Mathematical Sciences Institute,\\
      Australian National University, Canberra, ACT 0200,
      Australia.
}

\begin{abstract}
In this paper we continue the study of $Q$-operators in the
six-vertex model and its higher spin generalizations.
In \cite{Man14} we derived a new expression
for the higher spin $R$-matrix associated with the affine quantum algebra
$U_q(\widehat{sl(2)})$. Taking a special limit in this $R$-matrix
we obtained new formulas for  the $Q$-operators acting in
 the tensor product of 
representation spaces with arbitrary complex spin.

Here we use a different strategy and construct $Q$-operators as 
integral operators with factorized kernels based on the original
Baxter's method used in the solution of the eight-vertex model.
We compare
this approach with the method developed in \cite{Man14} and
find the explicit connection between two constructions.
We also discuss a reduction to the case of
finite-dimensional representations with (half-) integer spins.
\end{abstract}

\end{center}

\newpage

\nsection{Introduction}

In our previous paper \cite{Man14} we derived a new expression
for the $U_q(sl(2))$ $R$-matrix $R_{I,J}(\lambda)$
with a spectral parameter acting
in the tensor product of two highest weight representations with arbitrary
complex spins $I$ and $J$. The method we used was based on a 3D approach
developed in \cite{Bazhanov:2005as,Bazhanov:2008rd,MBS13}.
The result was surprisingly simple and represented in terms of
the basic hypergeometric function $\phifs$.

As an application of this new formula for the $R$-matrix we constructed 
the  $Q$-operators related to the $U_q(\widehat{sl(2)})$ algebra as  
special transfer matrices  acting in the tensor product of arbitrary highest 
weight representations. The idea of the construction of the $Q$-operator in 
terms of some special transfer matrices belongs
to Baxter \cite{Bax72}. It is a key element of his original solution
 of the 8-vertex model. For the simplest case of the six-vertex model 
 the quantum space is
built from 2-dimensional
highest weight representations of the $U_q(sl(2))$ algebra at every site of the lattice.

In 1997 Bazhanov, Lukyanov and Zamolodchikov suggested a new method to derive the
$Q$-operators related to the affine algebra $U_q(\widehat{sl(2)})$
\cite{BLZ97a,BLZ99a}. Based on the universal $R$-matrix theory
\cite{Drinfeld1985} they showed that the $Q$-operators can be constructed as  special
monodromy operators with the auxiliary space being an infinite-dimensional 
representation
of the $q$-oscillator algebra.
This method of construction of $Q$-operators has been actively used and extended
in \cite{BLZunpub,BJMS06,Boos2010, BLMS10,Staud13}.
However, the derivation of the local $Q$-operators  from the universal
$R$-matrix \cite{TK92} quickly becomes unbearable for higher spins.

Taking a special limit $I\to\infty$ \cite{Pronko99,YNZ05,Boos2013} 
in the $R$-matrix from \cite{Man14}
we generalized the construction of \cite{BLZ97a,BLZ99a}
to the case of arbitrary complex spin $s$. In the limit 
 $2s=I\to\mathbb{Z}_+$ the Verma module becomes reducible and
 contains an $(I+1)$-dimensional invariant subspace.
Our $Q$-operators allow a natural non-singular reduction to this
subspace.

Interestingly, there is a different approach to the construction of the $Q$-operators
based on the famous Baxter's ``propagation through the vertex'' method \cite{Bax82}.
Baxter's ideas were  developed 
by Bazhanov and Stroganov \cite{BS90} in their study of the six-vertex model
at roots of unity. They considered
fundamental $L$-operators \cite{Faddeev:1979} intertwined by the $R$-matrix of
the six-vertex model at the roots of unity $q^N=1$. In this case,
 the highest weight representation of the  $U_q(sl(2))$ algebra is replaced
with a cyclic representation. 
Based on factorization properties of the $U_q(sl(2))$ $L$-operator
all matrix elements of the
$Q$-operator can be explicitly calculated  as simple products involving only
a two-spin interaction.

The idea of ``factorized'' $Q$-operators was further developed by Pasquier and
Gaudin \cite{PG92} where
they constructed
the $Q$-operator for the Toda lattice in the form of an integral operator.
The integral $Q$-operator for the case of the XXX chain was first calculated
in \cite{Der99}.
Taking the limit $N\to\infty$ \cite{AuYang99} in the Bazhanov and Stroganov
construction \cite{BS90} one can recover the results of \cite{PG92} and \cite{Der99}.

The main difference of the  above approach from the original Baxter's method
is that the ``quantum'' representation space is infinite-dimensional.
It has the structure of a tensor product of Verma modules
with  the basis  chosen as multi-variable polynomials $p(x_1,\ldots,x_M)$,
where $M$ is the size of the system. 

Initially  the  $Q$-operator for the XXX spin chain for complex spins
was constructed as the integral operator \cite{Der99}. 
Later on a new operator approach was developed where the $Q$-operator becomes
a differential operator of infinite order \cite{CDKK12a}. 
This was generalized to the XXZ case when the $Q$-operator
is represented by a $q$-difference operator of the infinite order \cite{Der05,CDKK13}.

It is well known that in the XXZ case the solutions of $q$-difference equations
 can be fixed only up to certain periodic functions.
This can lead to potential problems when the Yang-Baxter equation is satisfied
only up to such periodic functions \cite{bazh11}. A possible resolution
in the XXZ case is to use a modular double \cite{Faddeev:1999}. The non-compact case and
applications of the $Q$-operators to Liouville theory are discussed in
\cite{Kas01,Bytsko:2006,Bytsko09}.

Recently there was a substantial progress in understanding of the structure of
a general $R$-matrix for the XYZ spin chain using elliptic modular double \cite{DS13}
where the elliptic beta integral plays an important role \cite{Spir2001}.
However, the integral form of the $Q$-operator for the XXZ spin chain was
still missing. In this paper we intend to fill this gap and present the
XXZ $Q$-operator as an integral operator acting in the space of polynomials.
In fact, we shall construct two such operators, one is based on the Askey-Roy
extension of the beta integral \cite{Askey86} and another is based
on the $q$-analogue of Barnes' first lemma \cite{Watson10}.

The next question is how these integral $Q$-operators are connected with the 
$Q$-operators derived in \cite{Man14}. We find the explicit relation
between them extending ideas of \cite{CDKK12b} for the XXX case, where
a connection between  the $Q$-operators of \cite{CDKK12a} and the
$Q$-operators of \cite{BLMS10} was found. 

The paper is organized as follows. In Section 2 we 
remind some basic facts about the XXZ chain
at arbitrary spin and explain the construction of the $Q$-operators
derived in \cite{Man14}. We also calculate the $Q$-operators
at some special value of the spectral parameter where they have a very
simple form. In Section 3 we introduce a polynomial representation 
of the $U_q(sl(2))$ algebra
and rewrite the transfer-matrix of the six-vertex model in the form of a finite order
$q$-difference operator in the space of $M$-variable polynomials.
In Section 4 we construct the XXZ $Q$-operator as the integral
operator acting in the space of polynomials and calculate its explicit
action. In Section 5 we prove a commutativity of the integral $Q$-operators
and the transfer-matrix at different values of the spectral parameters.
In Section 6 we find the explicit connection between the $Q$-operators
from Section 3 and the $Q$-operators constructed in \cite{Man14}.
Finally, in Conclusion we summarize all results and outline further
directions of investigation.

\nsection{The XXZ model at arbitrary spin}

In this section we remind some results from \cite{Man14}.
We start with the following $U_q(sl(2))$ $L$-operator
\beq
L(\lambda;\phi)=
\left(\begin{array}{ll}\ds
\phi^{-1}[\lambda q^{H/2}]&
\phi^{-1} [q]\,F \\
& \\
\phi [q]\,E &\phi[\lambda q^{-H/2}]\\
\end{array}\right)\label{xxz1}
\eeq
where
\beq
[x]=x-x^{-1},\label{xxz1a}
\eeq
$E$, $F$ and $H$ are the generators of the quantum algebra $U_q(sl(2))$
and $\phi$ is the horizontal field.

For any $I\in\mathbb{C}$ one can introduce an infinite-dimensional Verma module
$V_I^+$ with a basis $v_i$, $i\in\mathbb{Z}_{+}$. We define the
infinite-dimensional representation $\pi_I^+$ of $U_q(sl(2))$
by the following action on the module $V_I^+$
\beq
H v_i=(I-2i)v_i,\quad E v_i=\frac{[q^i]}{[q]}v_{i-1},\quad
F v_i=\frac{[q^{I-i}]}{[q]}v_{i+1}.
\label{xxz2}
\eeq
We notice that the $L$-operator (\ref{xxz1}) differs from the $L$-operator
used in \cite{Man14} by a change $\lam\to\lam q^{-1/2}$ and a simple similarity 
transformation in quantum space.
Namely, basis vectors $v_i$ used in this paper are related to basis vectors
$\tilde v_i$  in \cite{Man14} as
\beq
v_i=q^{-\frac{1}{2}i^2}\tilde v_i, \quad i=0,1,\ldots \label{xxz2a}
\eeq

When $I\in\mathbb{Z_+}$, the representation $\pi_I^+$ becomes reducible.
The vectors $v_i$, $i>I$ span an irreducible submodule of $V^+_I$ isomorphic
to $V^+_{-I-2}$ and one can introduce a finite-dimensional module
$V_I$ with the basis $\{v_0,\ldots,v_I\}$ isomorphic to
the quotient module $V_I^+/V_{-I-2}^+$. We denote the corresponding
finite-dimensional representation
as $\pi_I$.

For $I\in\mathbb{Z_+}$ the homogeneous transfer matrix
${\bf{T}}_{I}(\lam;\phi)$ with periodic boundary conditions
acting in the $(I+1)^M$-dimensional
quantum space $ W=\stackrel[i=1]{M}{\otimes}V_I$ is defined as
\beq
{\bf{T}}_{I}(\lambda;\phi)={\mbox{ Tr}}
\big[L_1(\lam;\phi)\otimes...\otimes L_M(\lam;\phi)\big]
\label{xxz3}
\eeq
where the trace is taken in the auxiliary space $\mathbb{C}^2$.

Due to a conservation law in the quantum space $W$
\beq
\sum_{k=1}^M i_k=\sum_{k=1}^M i'_k=l, \label{xxz4}
\eeq
the transfer matrix (\ref{xxz3}) has a block-diagonal form
\beq
\ds{{\bf T}}_{I}(\lambda;\phi)=
\ds\stackrel[l=0]{IM}{\ds\bigoplus}{{\bf T}}_{I}^{(l)}(\lambda;\phi).\label{xxz4a}
\eeq
We call the subspace in the quantum space $W$ with a fixed $l$ as the $l$-th sector
and denote it $W_l$.

Due to (\ref{xxz4}) one can introduce a simple equivalence transformation
in the quantum space and move all local fields $\phi$ to the left where
they combine into one boundary field $\phi^{-M\sigma_3}$. This is how twist is 
normally introduced into the XXZ model.
Here we prefer to use local fields $\phi$ because in this case we preserve
translational invariance of the spin chain which is obviously broken
in the case of a boundary field.

Let us notice that the direct sum expansion (\ref{xxz4}) is also true for 
$I\in\mathbb{C}$,
when the quantum space becomes infinite-dimensional, 
$W=\stackrel[i=1]{M}{\otimes}V_I^+$.
In this case the sum in (\ref{xxz4}) runs from zero to infinity,
but all blocks with a fixed $l$
are still finite-dimensional.

There are two $Q$-operators ${\bf Q}_\pm^{(I)}(\lam)$ which commute with
the transfer matrix ${\bf{T}}_{I}(\lambda;\phi)$ and among themselves
for different values of spectral parameters
\beq
[{\bf Q}_\pm^{(I)}(\lam),{\bf{T}}_{I}(\mu;\phi)]=
[{\bf Q}_+^{(I)}(\lam),{\bf Q}_-^{(I)}(\mu)]=
[{\bf Q}_\pm^{(I)}(\lam),{\bf Q}_\pm^{(I)}(\mu)]=0. \label{xxz5}
\eeq
They satisfy the famous Baxter $TQ$-relation
\beq
{\bf T}_{I}(\lambda;\phi){\bf Q}_\pm^{(I)}(\lambda)=
[\lambda/\zeta]^M {\bf Q}_\pm^{(I)}(q\lambda)+
[\lambda \zeta]^M {\bf Q}_\pm^{(I)}(q^{-1}\lambda),\label{xxz6}
\eeq
where we introduced a variable $\zeta$ for later convenience
\beq
\zeta=q^{I/2}. \label{xxz7}
\eeq

Let us first assume that $I\in\mathbb{Z}_+$.
In the  case $\phi\neq1$  the eigenvalues of ${\bf Q}_\pm^{(I)}(\lam)$
are polynomials in $\lambda$ and $\lam^{-1}$ up to a simple phase factor.
Namely, if we define two operators ${\bf A}^{(I)}_\pm(\lambda)$
\beq
{\bf Q}^{(I)}_\pm(\lambda)=e^{\pm i u h M}{\bf A}^{(I)}_\pm(\lambda)=
\lambda^{\pm h M}{\bf A}^{(I)}_\pm(\lambda), \label{xxz8}
\eeq
where
\beq
\lambda=e^{iu}, \quad \phi=q^h, \label{xxz9}
\eeq
then the eigenvalues ${\mathcal A}_\pm^{(I)}(\lambda)$ of
the operators
${\bf A}^{(I)}_\pm(\lambda)$ in the subspace $W_l$
have the following form
\beq
{\mathcal A}_+^{(I)}(\lambda)=\rho_+\prod_{k=1}^l [\lambda/\lambda_k^+],\quad
{\mathcal A}_-^{(I)}(\lambda)=
\rho_-\prod_{k=1}^{IM-l}[\lambda/\lambda_k^-].\label{xxz10}
\eeq
where $\lam_k^\pm$ are the solutions of the Bethe Ansatz equations.

The $TQ$-relation for the operators ${\bf A}^{(I)}_\pm(\lambda)$ takes the
form
\beq
{\bf T}_{I}(\lambda;\phi){\bf A}_\pm^{(I)}(\lambda)=
\phi^{\pm M}\,[\lambda/\zeta]^M {\bf A}_\pm^{(I)}(q\lambda)+
\phi^{\mp M}\,[\lambda \zeta]^M
{\bf A}_\pm^{(I)}(q^{-1}\lambda).\label{xxz10a}
\eeq

Let us notice that we could change a $\lam$-dependent normalization
of the operators ${\bf A}^{(I)}_\pm(\lambda)$
in such a way that their eigenvalues would become polynomials in $\lam^2$.
Such a change of normalization will result in additional factors $q^{\pm l}$
in the RHS of (\ref{xxz10a}).

Operators ${\bf A}^{(I)}_\pm(\lambda)$ satisfy the Wronskian relation
\beq
\phi^M {\bf A}_+^{(I)}(q\lambda )
{\bf A}_-^{(I)}(\lambda )-\phi^{-M}{\bf A}_-^{(I)}(q\lambda )
{\bf A}_+^{(I)}(\lambda)={\bf Wr}(\phi)
\lam^{IM}(\lam^{-2}q^{-I};q^2)_I^M,\label{xxz10b}
\eeq
where we defined the $q$-Pochhammer symbol
\beq
(x;q)_n=\prod_{k=0}^{n-1}(1-x q^k).\label{xxz10c}
\eeq

The Wronskian ${\bf Wr}(\phi)$ was calculated in \cite{Man14}
\beq
\text{\bf Wr}(\phi)
=(-1)^{I M}\phi^M q^{l-I M}(1-\phi^{2M} q^{2l-I M}) {\boldsymbol I}. \label{xxz10d}
\eeq

Operators ${\bf A}^{(I)}_\pm(\lambda)$ were constructed  
in \cite{Man14}
as special transfer matrices acting in the  subspace $W_l$
\beq
{\bf A}^{(I)}_\pm(\lambda)=(1-\phi^{2M}q^{2l-I M})\times
\underset{{\,\mathcal{F}_q}}{{\mbox{Tr}}}\,\{
\underbrace{
A^{(I)}_\pm(\lambda)\otimes\ldots\otimes 
A^{(I)}_\pm(\lambda)}_\text{$M$ times}\},\label{xxz11}
\eeq
where the trace is calculated over the  infinite-dimensional Fock space
$\mathcal{F}_q$, spanned by a set of vectors
$|n\rangle$, $n=0,1,2,\ldots,\infty$.  We always choose the field 
$\phi\in\mathbb{C}$ in (\ref{xxz11}) to ensure
a convergency of the geometric series in (\ref{xxz11})
and then analytically continue
to all values of $\phi$.

Local $L$-operators acting in the tensor product $\mathcal{F}_q\otimes
V_I$  have the form
\beq
\begin{split}
[A^{(I)}_+(\lambda)]_{n,i}^{n',i'}=
\delta_{i+n',i'+n}&\,{\phi^{-2n}}(-1)^{i+i'}\lambda^{-i}
q^{\frac{1}{2}i(i+1)-\frac{1}{2}i'(i'+1)+i(I+i')+n(I-i-i')}\times \\
&\times\frac{(q^{2};q^2)_{n'}}
{(q^2;q^2)_n(q^2;q^2)_i\, }\,
{\prescript{}{3}{\overline\phi}}_2\left(\left.\begin{array}{l}
q^{-2i};q^{-2i'},\lambda^2q^{-I}\\
q^{-2I},q^{2(1+n-i)}\end{array}\right|q^2,q^2\right)\label{xxz12}
\end{split}
\eeq
and
\beq
\begin{split}
[A^{(I)}_-(\lambda)]_{n,i}^{n',i'}=&\delta_{i+n,i'+n'}\,{\phi^{2n}}\lambda^{i-I}
q^{-\frac{1}{2}i(i-1)+\frac{1}{2}i'(i'-1)+i(I+i')+n(I-i-i')}\times \\
&\times\frac{(\lambda^2q^{-I+2(i'-n)};q^2)_{I-i-i'}}
{(q^2;q^2)_i\,\,}\,
{\prescript{}{3}{\overline\phi}}_2\left(\left.\begin{array}{l}
q^{-2i};q^{-2i'},\lambda^2q^{-I}\\
q^{-2I},q^{2(1+n-i')}\end{array}\right|q^2,q^2\right),
\end{split}\label{xxz13}
\eeq
where we  defined  a {\it regularized} terminating basic hypergeometric series $\phir$
 as
\beq
\ds
\phir\left(\left.\begin{array}
{l}q^{-n};a_1,\ldots,a_r\\
\phantom{q^{-n},}b_1,\ldots,b_r\end{array}\right|q,z\right)=
\sum_{k=0}^n z^k\,\frac{(q^{-n};q)_k}{(q;q)_k}\prod_{s=1}^r(a_s;q)_k (b_sq^k;q)_{n-k}\,.
\label{xxz14}
\eeq
Such a regularization is necessary, since the parameter $q^{2(1+n-i)}$ in
(\ref{xxz12}) may become
equal to  $q^{-2m}$, $m=0,1,\ldots$ and the standard
basic hypergeometric series is not defined at these points.

It was shown in \cite{Man14} that two $L$-operators (\ref{xxz12}-\ref{xxz13})
are related by the following transformation for integer $I\in\mathbb{Z}_+$,
$0\leq i,i'\leq I$
\beq
[A^{(I)}_-(\lambda)]_{n,i}^{n',i'}=
[A^{(I)}_+(\lambda)]_{n,I-i}^{n',I-i'}|_{\phi\to\phi^{-1}}.\label{xxz15}
\eeq
In fact, we could use (\ref{xxz15}) to define the second $Q$-operator
${\bf A}^{(I)}_-(\lambda)$ in terms of the first one, ${\bf A}^{(I)}_+(\lambda)$.

Due to the presence of $\delta$-functions in
(\ref{xxz12}-\ref{xxz13})
both $Q$-operators act invariantly
in  subspaces $W_l$ similar to the transfer-matrix ${\bf T}_{I}(\lambda;\phi)$.

The choice of a normalization factor in (\ref{xxz11}) leads to simple asymptotics
of both operators ${\bf A}^{(I)}_\pm(\lambda)$ at $\lam\to\infty$
\begin{align}
&{\bf A}_+^{(I)}(\lambda)|_{\lambda\to\infty} =
-(-\lambda)^l\phi^{2M}q^{l-IM}({\boldsymbol I}+O(\lambda^{-2})),
\nonumber\\
&{\bf A}_-^{(I)}(\lambda)|_{\lambda\to\infty} =\phantom{-}
(-\lambda)^{I M-l}q^{l-IM}({\boldsymbol I}+O(\lambda^{-2})).\label{xxz16}
\end{align}

The $L$-operator (\ref{xxz12}) can be continued to arbitrary $I\in\mathbb{C}$,
since its matrix elements are polynomials in $\zeta^2$  where $\zeta$ was defined in
(\ref{xxz7}). This defines
the $Q$-operator ${\bf A}_+^{(I)}(\lam)$ for any $I\in\mathbb{C}$.
The continuation of the second $Q$-operator to non-integer values of $I$ is 
more problematic.
The $L$-operator (\ref{xxz13}) has a pre-factor which contains
infinitely many zeros and poles in $\zeta^2$ for non-integer $I$. If we
multiply (\ref{xxz13})
by a simple meromorphic function independent of indices
\beq
(\lam^{-2}q^{-I};q^2)_I^{-1}=\frac{(\lam^{-2}q^{I};q^2)_\infty}
{(\lam^{-2}q^{-I};q^2)_\infty},\label{xxz17}
\eeq
then the matrix elements of the $L$-operator (\ref{xxz13}) contain
a ratio of two $q$-Pochhammer symbols.
Since  such a ratio is no longer a polynomial in $\zeta^n$ and $q^n$,
a calculation of the trace over the auxiliary Fock space becomes a  nontrivial
problem.
Let us also notice that a restriction of both $L$-operators from generic 
to positive integer values of $I$ is non-singular and 
the corresponding transfer-matrices
act invariantly in the finite-dimensional quantum space $V_I$.

Both $L$-operators (\ref{xxz12}-\ref{xxz13}) simplify significantly for 
two special values of the spectral parameter, $\lam=\zeta^{\pm1}$.
At $\lam=\zeta$ (\ref{xxz12}) reduces to
\begin{align}
[A^{(I)}_+(\zeta)]_{n,i}^{n',i'}=
\delta_{i+n',i'+n}&(-1)^{i'}q^{ii'+\frac{1}{2}i(I+3-i)-\frac{1}{2}i'(i'+1)
}\frac{(q^{-2I};q^2)_i}{(q^2;q^2)_i}\times\nonumber\\
&\times\frac{(q^2;q^2)_{n'}}{(q^2;q^2)_n}
\phi^{-2n}q^{n(I+i-i')}
(q^{-2n};q^2)_i. \label{xxz18}
\end{align}
Now we can expand the last $q$-Pochhammer symbol in (\ref{xxz18}) in series 
in $q^{-2n}$, take the tensor product of $M$ copies of the $L$-operator and
calculate the trace over the Fock space in (\ref{xxz11}). The result can be
written as an $M$-fold sum. One can remove one summation using
the following simple identity
\beq
\sum_{k=0}^i\frac{(q^{-2i};q^2)_k}{(q^2;q^2)_k}
\frac{q^{2i k}}{1-x q^{-2k}}=-x^{-1}\frac{(q^2;q^2)_i}{(x^{-1};q^2)_{i+1}},
\quad i\in\mathbb{Z}_+,\quad x\in \mathbb{C}. \label{xxz19}
\eeq

The final result for the action
of the operator ${\bf A}_+^{(I)}(\zeta)$ in the quantum subspace
$W_l$
can be written in the following neat form
\begin{align}
\left[{\bf  A}_+^{(I)}(\zeta)\right]_{i_1,\ldots,i_M}^{i_1',\ldots,i_M'}&=
(-1)^{l+1}(1-\phi^{2M}q^{2l-I M})q^l\zeta^{l}
\prod_{k=1}^M\frac{(\zeta^{-4};q^2)_{i_k}}{(q^2;q^2)_{i_k}}
(\phi/\zeta)^{2+2k(i_k'-i_k)}\times
\nonumber\\
&\times\sum_{s=0}^{l-i_1}
\frac{(q^2;q^2)_{i_1}}{(\phi^{2M}\zeta^{-2M}q^{2s};q^2)_{i_1+1}}
\frac{1}{s!}\frac{d^s}{dz^s}
\prod_{m=2}^M(z q^{2+2\sum_{l=1}^{m-1}(i_l-i_l')};q^2)_{i_m}\Bigg|_{z=0}.\label{xxz20}
\end{align}

In (\ref{xxz20}) we introduced a dummy variable $z$ to uncouple remaining
$M-1$ summations and converted back to the product of $M-1$ $q$-Pochhammer symbols.
It is quite surprising that there exist such
a simple expression for the transfer-matrix (\ref{xxz11}) at the particular
value of $\lam=\zeta$. We also notice that a derivation of the result (\ref{xxz20})
remains valid for any $I\in\mathbb{C}$. In Section 6 we will show that this formula
allows us to explicitly compute matrix elements
of the transfer-matrix (\ref{xxz11}) for arbitrary values of $\lam$.

At the point $\lam=\zeta^{-1}$ the hypergeometric function $\phit$
reduces to $\phis$ of the argument $q^2$ and can be calculated using
the $q$-Vandermonde sum. As a result we obtain the expression for the $L$-operator
(\ref{xxz12}) similar to (\ref{xxz18}) and again can calculate the
transfer-matrix in a closed form. We shall not use this expression and
leave it as an exercise for the reader.

In the case $I\in\mathbb{Z}_+$ the second transfer-matrix
${\bf A}_-^{(I)}(\zeta)$ can be obtained from (\ref{xxz20}) using a symmetry
(\ref{xxz15}). The difficult case $I\in\mathbb{C}$ will be discussed elsewhere.

\nsection{Polynomial representation}

There is another representation of the quantum algebra $U_q(sl(2))$
on the space of polynomials which we also use in this paper.

Let us introduce the polynomial ring $K[x]$ in variable $x$
over the field $\mathbb{C}$
and its multi-variable generalization
$K_M[X]$, $X\equiv\{x_1,\ldots,x_M\}$
and identify basis
vectors
in $W$ with monomials in $K_M[X]$
\beq
v_{i_1,\ldots,i_M}\equiv
v_{i_1}\otimes\ldots\otimes v_{i_M} \sim x_1^{i_1}\ldots x_M^{i_M}.\label{pol1}
\eeq

The ring $K_M[X]$ has a gradation
\beq
K_M[X]=\bigoplus_{l\in \mathbb{Z_+}} K^{(l)}_M[X], \label{pol2}
\eeq
where $K_M^{(l)}[X]$ is generated by monomials in $M$ variables of 
the total degree $l$.
There is an obvious isomorphism between $W_l$ and $K_M^{(l)}[X]$.

In this paper we consider only periodic boundary conditions
and always imply periodicity  $M+1\equiv 1$ and all indices run from $1$ to
$M$ $(\mod M)$, i.e. $x_0=x_M$, etc.

Let us introduce a set of operators ${\mathcal X}_i$ and 
${\mathcal D}_i$, $i=1,\ldots,M$
acting in $K_M[X]$ as
\beq
{\mathcal X}_i p(x_1,\ldots,x_M)=x_i p(x_1,\ldots, x_M),\quad
{\mathcal D}_i\, p(x_1,\ldots,x_M)=p(x_1,\ldots, q x_i,\ldots, x_M). \label{pol3}
\eeq
The $U_q(sl(2))$ generators $H$, $E$, $F$ in $K[x]$ can be realized in 
terms of one pair of operators $\ml X$, $\ml D$
\beq
q^H=\zeta^2 \ml D^{-2},\quad E=\ml X^{-1}\frac{[\ml D]}{[q]},
\quad F=\ml X\frac{[\zeta^2 \ml D^{-1}]}{[q]}, \label{pol4}
\eeq
where $\zeta$ was defined in (\ref{xxz7}).

The $L$-operator takes the form
\beq
L(\lambda;\phi)=
\left(\begin{array}{ll}\ds
\phi^{-1}[\lam z\ml D^{-1}]&
\phi^{-1}\ml X [z^2 \ml D^{-1}] \\
& \\
\phi \ml X^{-1} [\ml D] &\phi[\lambda z^{-1}\ml D]\\
\end{array}\right).\label{pol6a}
\eeq

Again we can define the
transfer-matrix with periodic boundary conditions
\beq
{\bf{T}}_{I}(\lambda;\phi)={\mbox{ Tr}}
\big[L_1(\lam;\phi)\otimes...\otimes L_M(\lam;\phi)\big]
\label{pol7}
\eeq
which is a
 $q$-difference operator in $M$ variables acting on polynomials
$p(x_1,\ldots,x_M)\in K_M[X]$.

The $TQ$-relation becomes the operator relation acting in $K_M[X]$
\beq
{\bf T}_I(\lam;\phi){\bf A}_\pm(\lam)=\phi^{\pm M} [\lam/\zeta]^M
 {\bf A}_\pm(q\lam)+\phi^{\mp M}
[\lam \zeta]^M{\bf A}_\pm(q^{-1}\lam). \label{pol8}
\eeq

Although the transfer-matrix (\ref{pol7}) is a finite order
$q$-difference operator,
$Q$-operators will be, in general, integral operators.
Our goal is to construct $Q$-operators as special integral operators
with factorized kernels acting in $K_M[X]$.

\nsection{Integral $Q$-operator}

In this section we construct the XXZ $Q$-operator as an integral
operator with a factorized kernel. First we consider the case $I\in\mathbb{C}$
and discuss a restriction to the finite-dimensional case later.

We will denote the $Q$-operator in this section  as ${\bf Q}_f(\lam)$.
It has polynomial eigenfunctions and should not be confused
with the $Q$-operators ${\bf Q}_\pm(\lam)$ from Section 2.

First we assume that the Q-operator can be represented as
an integral operator with the kernel ${Q}_\lam({\bf x}|{\bf y})$
\beq
[{\bf Q}_f(\lam)p]({\bf x})=
\ds\oint\limits_{\ml C} \frac{dy_1}{y_1}\ldots\frac{dy_M}{y_M}\>
Q_\lam({\bf x}|{\bf
y}) p({\bf y}), \label{pol9}
\eeq
where ${\bf x}\equiv\{x_1,\ldots,x_M\}$, ${\bf y}\equiv\{y_1,\ldots,y_M\}$ and
the integration contour $\ml C$ is a proper deformation of the unit circle.

In general, one can expect that  the kernel $Q_\lam({\bf x}|{\bf
y})$ will have a non-local structure, i.e. involve
functions depending simultaneously on all variables ${\bf x}$ and
${\bf y}$. Our goal is to construct
a factorized
kernel $Q_\lam({\bf x}|{\bf y})$ which can be represented as a product
of factors depending on pairs of local variables.

It is well known that this can be achieved by
using Baxter's method of ``pair-propagation through a vertex''.
As a result the $TQ$-relation reduces to a set of difference equations
in variables ${ x_i}$ for the kernel $Q_\lam({\bf x}|{\bf y})$.
However, in this approach we can not fix a dependence of
the kernel on the integration variables ${y_i}$.
This difficulty can be resolved by
applying Baxter's method to a ``conjugated'' $L$-operator.
Such approach will allow us first to determine
a dependence of the kernel
on ${y}$-variables in (\ref{pol9}). Then a dependence on
${x}$-variables can be fixed by a proper normalization of the $Q$-operator. 
In original Baxter's notations
it is equivalent to a construction of the operator $Q_L(\lam)$.

Let us consider a finite-order $q$-difference operator $M$ and
define its conjugate $M^*$
by
\beq
\oint\limits_{\ml C} \frac{dy}{y} f(y) [M g](y)=\oint\limits_{\ml C}
\frac{dy}{y}[M^* f](y) g(y),\label{pol10}
\eeq
where functions $f(x)$ and $g(x)$ are properly defined to ensure a convergence
of the integral. Here we also assume that the integration contour is invariant
under the change of variables $y\to q^{\pm 1} y$.

In particular, we have
\beq
\ml X^*=\ml X,\quad \ml D^*=\ml D^{-1}.\label{pol11}
\eeq

Using (\ref{pol11}) one can construct the $L$-operator conjugated
to (\ref{pol6a}) in the sense (\ref{pol10})
\beq
L^*(\lambda;\phi)=
\left(\begin{array}{ll}\ds
\phi^{-1}[\lam z\ml D]&
\phi^{-1}\ml X [q z^2 \ml D] \\
& \\
\phi \ml X^{-1} [q \ml D^{-1}] &\phi[\lambda z^{-1}\ml D^{-1}]\\
\end{array}\right)\label{pol12}
\eeq
and define the associated transfer-matrix
\beq
{\bf{T}}_{I}^*(\lambda;\phi)={\mbox{ Tr}}
\big[L_1^*(\lam;\phi)\otimes...\otimes L_M^*(\lam;\phi)\big]
\label{pol13}
\eeq

Now let us introduce a set of local transformations $U_i$ in the auxiliary
space $\mathbb{C}^2$
\beq \tilde
L^*_i(\lam)= U_{i-1}\>L^*_i(\lam)\>U^{-1}_{i},\quad
U_i=\left(\begin{array}{cc} 1 & \alpha_i \\ 0& 1 \end{array}\right),
\quad i=1,\ldots, M,\label{pol14}
\eeq
where we defined a set of arbitrary complex parameters $\alpha_i$,
$i=1,\ldots,M$. Obviously the transfer matrix (\ref{pol13}) is not
affected by this
transformation of the $L$-operators.

According to the Baxter's method we now construct a set of functions $f(y_i)$
satisfying
\beq [\tilde
L^*_i(\lam)f]({y_i})= \left(\begin{array}{cc} * & 0
\\ * &* \end{array}\right). \label{pol15}
\eeq
From (\ref{pol15}) it is easy to obtain a recurrence relation
for the function $f(y_i)$:
\beq
\frac{f(q^2y_i)}{f(y_i)}= \ds q^2
\frac{\ds\left(1-\frac{y_i}{\lam\zeta \alpha_{i-1}\phi^2}\right)
\left(1-\frac{\lam y_i}{\zeta\alpha_i }\right)}
{\ds\left(1-\frac{q^2\lam \zeta y_i }{\alpha_{i-1}\phi^2}\right)
\left(1-\frac{q^2 \zeta y_i }{\lam \alpha_i}\right)
}. \label{pol16}
\eeq
Let us notice that the  RHS of (\ref{pol16}) factorizes into the product
of four factors even in the case of arbitrary horizontal field $\phi$.
As a result the horizontal field can be nontrivially introduced into
the periodic ``saw''-like structure of the kernel of the $Q$-operator.
To our knowledge this was not known before.

There are infinitely many solutions to the difference equation
(\ref{pol16}). They all differ by functions which are periodic
in $q^2$, i.e. for any two solutions $f_1(y)$, $f_2(y)$ of (\ref{pol16})
we have
\beq
f_1(y)=g(y)f_2(y),\quad g(q^2y)=g(y). \label{pol17}
\eeq
The choice of the function $g(y)$ is determined by the condition that
$Q$-operators should commute at different values of spectral parameters
as {\it integral} operators. This can be also reformulated as  two statements
which are more appropriate in our context.
First, we show that the $Q$-operator maps polynomials to polynomials, i.e.
$K_M[X]$ to $K_M[X]$. Second, we require that $Q$-operators commute
on polynomials. Since polynomial functions are dense in the space
of continuous complex-valued functions on a finite interval,
it would imply a commutativity of $Q$-operators as integral operators.

In this paper we are going to introduce two types of integral operators.
The first one is of the type (\ref{pol9}) and based
on Askey and Roy extension of the beta integral \cite{Askey86} (see
also (4.11.2) in \cite{Gasper})
\beq
\frac{1}{2\pi i}
\oint\limits_{\ml C}
\frac{\ds\left(\frac{\rho y}{d},\frac{qd}{\rho y},\frac{\rho c}{y},
\frac{qy}{\rho c};q\right)_\infty}
{\ds\left(ay,by,\frac{c}{y},\frac{d}{y};q\right)_\infty}\,\frac{dy}{y}=
\frac{\ds\left(abcd,\rho,\frac{q}{\rho},\frac{\rho c}{d},\frac{q d}{\rho c};q
\right)_\infty}
{\ds \left(ac,ad,bc,bd,q;q\right)_\infty}, \label{pol18}
\eeq
where $ac,ad,bc,bd\neq q^{-n}$, $n=0,1,2,\ldots$, $\rho cd\neq0$ and
the contour $\ml C$ is a deformation of the unit circle such that
the zeros of $(ay,by;q)_\infty$ lie outside the contour and the origin
and zeros of $(c/y,d/y;q)_\infty$ lie inside the contour.

Motivated by (\ref{pol18}) we choose a solution of (\ref{pol16}) as
\beq
f(y_i)=\mu(\alpha_{i-1},\alpha_i)\frac{1}{2\pi i}
\frac{\left(\ds\frac{\lam\zeta \rho_i y_i}{\alpha_{i-1}\phi^2},
\frac{q^2\alpha_{i-1}\phi^2}{\lam\zeta\rho_i y_i},
\frac{\lam\rho_i\alpha_i}{\zeta y_i},
\frac{q^2\zeta y_i}{\lam\rho_i\alpha_i};q^2\right)_\infty}
{\ds \left(\frac{y_i}{\lam\zeta\alpha_{i-1}\phi^2},
\frac{\alpha_{i-1}\phi^2}{\lam\zeta y_i},
\frac{\lam y_i}{\zeta \alpha_i},
\frac{\lam \alpha_i}{\zeta y_i};q^2\right)_\infty},\label{pol19}
\eeq
where $\mu(\alpha_{i-1},\alpha_i)$ are some normalization factors
and $\rho_i$ are the parameters at our disposal.
It is easy to check that (\ref{pol19}) satisfies (\ref{pol16}).

Now we are ready to define our $Q$-operator. The product of $M$ functions
(\ref{pol19}) contains $M$ arbitrary parameters $\alpha_i$ and we identify
them with $x_i$ in (\ref{pol9}). Then we fix the factors
$\mu(x_{i-1},x_i)$ in such a way that the $Q$-operator has the following
normalization
\beq
{\bf Q}_f(\lam)\cdot {\bf 1}={\bf 1}. \label{pol20}
\eeq
Choosing $\rho_i= x_{i-1}\lam^{-1}\phi^2$ we obtain the action of the $Q$-operator
\begin{align}
[{\bf Q}_f(\lam)g]&(x_1,\ldots,x_M)=\frac{1}{(2\pi i)^M}
\prod\limits_{k=1}^M
\frac{\ds\left(\frac{\lam^2}{\zeta^2},\frac{1}{\lam^2\zeta^2},
\frac{x_k}{\zeta^2 x_{k-1}\phi^2},
\frac{x_{k-1}\phi^2}{\zeta^2 x_k},q^2;q^2\right)_\infty}
{\ds\left(\zeta^{-4},
\frac{x_{k-1}\phi^2}{\lam},
\frac{q^2\lam}{x_{k-1}\phi^2},
{\lam x_k},\frac{q^2}{\lam x_k};q^2\right)_\infty}\times\nonumber\\
&\times\oint\limits_{\ml C_1}\frac{dy_1}{y_1}\ldots
\oint\limits_{\ml C_M}\frac{dy_M}{y_M}\,g(y_1,\ldots,y_M)
\prod\limits_{k=1}^M
\frac{\left(\ds{\zeta  y_k},
\frac{q^2}{\zeta y_k},
\frac{x_{k-1}x_k\phi^2}{\zeta y_k},
\frac{q^2\zeta y_k}{x_{k-1}x_k\phi^2};q^2\right)_\infty}
{\ds \left(\frac{y_k}{\lam\zeta x_{k-1}\phi^2},
\frac{x_{k-1}\phi^2}{\lam\zeta y_k},
\frac{\lam y_k}{\zeta x_k},
\frac{\lam x_k}{\zeta y_k};q^2\right)_\infty}\,
,\label{pol21}
\end{align}
where contours $\ml C_1,\ldots,\ml C_M$ are defined similar to
the contour in (\ref{pol18}).
Using  the Askey-Roy integral (\ref{pol18}) it is easy to check
that the $Q$-operator defined by (\ref{pol21}) satisfies
the normalization condition (\ref{pol20}).

Having the explicit form of the kernel $Q_\lam({\bf x}|{\bf y})$ we can
calculate the diagonal matrix elements in (\ref{pol15}).
Using standard arguments we come to the ``left'' $TQ$-relation
acting in $K_M[X]$
\beq
{\bf Q}_f(\lambda){\bf T}_{I}(\lambda;\phi)=
\phi^{ M}\,[\lambda/\zeta]^M {\bf Q}_f(q\lambda)+
\phi^{- M}\,[\lambda \zeta]^M
{\bf Q}_f(q^{-1}\lambda).\label{pol21a}
\eeq
Let us notice that if we only use the relation
(\ref{pol16}), then the diagonal matrix elements in (\ref{pol15}) will contain
``wrong'' factors $[\lam/q\zeta]$ and $[q\lam\zeta]$ instead of
the ``right'' ones $[\lam/\zeta]$ and $[\lam\zeta]$. However,
the kernel of the $Q$-operator in (\ref{pol21}) contains
a pre-factor which we obtained
from the normalization condition on the $Q$-operator.
This pre-factor will generate the  missing factors
$[\lam/\zeta]$ and $[\lam\zeta]$
and cancel the ``wrong'' ones.

The next step is to show that the integral operator defined by (\ref{pol21})
maps polynomials in $M$ variables into polynomials in $M$ variables.
This is achieved by choosing a proper basis.
Let us fix real parameters $x_i$, $i=1,\ldots,M$ and
consider a family of polynomials
in $y_i$, $i=1,\ldots,M$
\beq
P_{\{n\},\{m\}}(y_1,\ldots, y_M)=\prod_{i=1}^M p_{n_i,m_i}(y_i),\label{pol22}
\eeq
where
\beq
p_{n_i,m_i}(y_i)=\left(\frac{y_i}{\lam\zeta x_{i-1}\phi^2};q^2\right)_{n_i}
\left(\frac{\lam y_i}{\zeta x_i};q^2\right)_{m_i} \label{pol23}
\eeq
and $\{n\}\equiv\{n_1,\ldots,n_M\}$, $\{m\}\equiv\{m_1,\ldots,m_M\}$
are two arbitrary sequences of positive integers.
Obviously, polynomials (\ref{pol22}) form a basis in $K_M[Y]$. In fact,
it is sufficient to choose all $n_i=0$ and consider only nonzero $m_i$'s.

Using a simple formula
\beq
(x;q^2)_n=\frac{(x;q^2)_\infty}{(xq^{2n};q^2)_\infty}
\eeq
and (\ref{pol18}) one can easily check that the $Q$-operator (\ref{pol21})
maps polynomials $p_{n_i,m_i}(y_i)$ into polynomials in $x_{i-1}/x_i$ and
$x_{i}/x_{i-1}$
\beq {\bf Q}_f(\lam): p_{n_i,m_i}(y_i)\to
\frac{\left(\ds \frac{x_i}{\zeta^2x_{i-1}\phi^2},\frac{1}{\lam^2\zeta^2};q^2\right)_{n_i}
\left(\ds \frac{x_{i-1}\phi^2}{\zeta^2 x_i},\frac{\lam^2}{\zeta^2};q^2\right)_{m_i}}
{(\zeta^{-4};q^2)_{n_i+m_i}}.\label{pol24}
\eeq
Since the kernel of the Q-operator is factorized, it maps
a product of polynomials $p_{n_i,m_i}(y_i)$ for different $i$ into the
product of RHS's in (\ref{pol24}). This completely describes the
action of the Q-operator (\ref{pol21}) on the basis of polynomials in $y_i$,
$i=1,\ldots,M$.

To make this action  explicit (the LHS of (\ref{pol24})
involves a dependence on the variables $x_i$'s) let us use the
following simple formula
\beq
x^m=\sum_{k=0}^m\frac{(q^{-m};q)_k}{(q;q)_k}q^k(x;q)_k. \label{pol25}
\eeq

It allows us to expand the powers $x^m$ in terms of q-binomials.
Combining it with the action (\ref{pol24}) and setting $n_i=0$ we
obtain
\beq {\bf Q}_f(\lam): y_i^m\to
\left(\frac{\zeta x_i}{\lam}\right)^m\sum_{k=0}^m
\frac{\left(\ds q^{-2m},\frac{\lam^2}{\zeta^2},
\frac{x_{i-1}\phi^2}{x_i\zeta^2};q^2\right)_k}
{(q^2,\zeta^{-4};q^2)_k}q^{2k}
\label{pol26}
\eeq
We can simplify it even further by expanding the last $q$-Pochhammer symbol
in the numerator in series and performing a summation in $k$. The result
reads
\beq
{\bf Q}_f(\lam): y_i^m\to\left(\frac{\lam}{\zeta}\right)^m
\frac{(q^2;q^2)_m}{(\zeta^{-4};q^2)_m}
\sum_{k=0}^m x_{i-1}^k x_i^{m-k}
\left(\frac{\phi^2}{\lam^2}\right)^k
\frac{(\lam^2\zeta^{-2};q^2)_k(\lam^{-2}\zeta^{-2};q^2)_{m-k}}
{(q^2;q^2)_k(q^2;q^2)_{m-k}}. \label{pol27}
\eeq
This formula clearly demonstrates that the integral operator defined by
the action (\ref{pol21}) maps polynomials in $M$ variables into polynomials
in $M$ variables. Moreover, the total degree of the polynomials is conserved,
so it acts invariantly in the subspace $K^{(l)}_M[X]$ of homogeneous polynomials
of the degree $l$.

Let us notice that if we set $\lam=\zeta$, then the action
(\ref{pol27}) is trivial and the $Q$-operator ${\bf Q}_f(\lam)$
simply becomes the identity operator
\beq
{\bf Q}_f(\lam)={\bf I}. \label{pol27a}
\eeq

Multiplying (\ref{pol27}) by
\beq
\zeta^{2m}\phi^{-m}\mu_i^m \frac{(\zeta^{-4};q^2)_m}
{(q^2;q^2)_m} \label{pol28}
\eeq
 and summing over $m$ from $0$ to $+\infty$ we derive the action
of the $Q$-operator on a generating function of polynomials in $y_i$.
Taking the product over $i=1,\ldots,M$ we obtain
\beq
{\bf Q}_f(\lam) \cdot
\prod_{i=1}^M\frac{(\zeta^{-2}\phi^{-1}\mu_i x_i;q^2)_\infty}
{(\zeta^{2}\phi^{-1}\mu_i x_i;q^2)_\infty}=
\prod_{i=1}^M
\frac{(\lam\zeta^{-1}\phi\mu_i x_{i-1};q^2)_\infty}
{(\zeta\lam^{-1}\phi\mu_i x_{i-1};q^2)_\infty}
\frac{((\lam\zeta\phi)^{-1}\mu_i x_{i};q^2)_\infty}
{(\lam\zeta\phi^{-1}\mu_i x_{i};q^2)_\infty},\label{pol29}
\eeq
where $\mu_i$ are arbitrary complex parameters and we imply
a periodicity in $M$, i.e. $x_0\equiv x_M$.
This formula describes the action of the $Q$-operator on arbitrary
polynomials in $M$ variables and coincides with
the formula (3.25) of \cite{CDKK13} at $\phi=1$. However,
the method we used is completely different from the method of \cite{CDKK13}.
Our $Q$-operator is a well defined integral operator and the $Q$-operator
constructed in \cite{CDKK13} is the difference operator of infinite order.

One can construct another integral operator which has the same polynomial
action (\ref{pol27}). The kernel of this integral operator comes
from the $q$-analogue of Barnes' first lemma discovered by Watson \cite{Watson10}
\begin{align}
\frac{1}{2\pi i} \int\limits_{-i\infty}^{i\infty}&
\frac{(q^{1-c+s},q^{1-d+s};q)_\infty}{(q^{a+s},q^{b+s};q)_\infty}
\frac{\pi q^s ds}{\sin \pi(c-s)\sin\pi(d-s)}=\nonumber\\
&=\frac{q^c}{\sin\pi(c-d)}\frac{(q,q^{1+c-d},q^{d-c},q^{a+b+c+d};q)_\infty}
{(q^{a+c},q^{a+d},q^{b+c},q^{b+d};q)_\infty}, \label{pol30}
\end{align}
where the contour of integration runs from $-i\infty$ to $i\infty$
and separates increasing and decreasing sequences of poles. To use (\ref{pol30})
for the construction of the $Q$-operator we need to make a change of variables.
Let us define two new sets of variables $t_i$, $s_i$ as
\beq
x_i=q^{2t_i},\quad y_i=q^{2s_i},\quad i=1,\ldots,M. \label{pol31}
\eeq
Then we can write another solution of (\ref{pol16})
\beq
f(y_i)=g(\lam,y_i)y_i\frac{\ds\left(\frac{q^2\lam
y_i\zeta}{x_{i-1}\phi^2},\frac{q^2y_i\zeta}{\lam x_i};q^2\right)_\infty}
{\ds\left(\frac{y_i}{\lam\zeta x_{i-1}\phi^2},\frac{\lam
y_i}{x_i \zeta};q^2\right)_\infty}
\eeq
where the function $g(\lam,y)$ satisfies
\beq g(q\lam,qy)=g(\lam,q^2y)=g(q^2\lam,y)=g(\lam,y). \label{pol32}
\eeq
It is easy to see that such function does not affect the $TQ$-relation
because it is periodic in $q^2$ and shifts in $\lam$ and $y$ always
enter in pairs. The Barnes' first lemma  (\ref{pol30}) suggests
to choose the function $g(\lam,y)$ as
\beq
g(\lam, y_i)=
\frac{\pi q^{2s_i}}{\ds\sin\pi(t_i-s_i+v-{I}/{4})\sin\pi(t_{i-1}-s_i+h-v-{I}/{4})},
\label{pol33}
\eeq
where
\beq
\lam=q^{2v},\quad \phi=q^h. \label{pol34}
\eeq
It is easy to check that (\ref{pol33}) satisfies (\ref{pol32}).

Using (\ref{pol30}) we can again normalize the $Q$-operator such that
it satisfies (\ref{pol20}). Then we come to a different
integral representation for the $Q$-operator
\begin{align}
&{\ds [{\bf Q}_f(\lam) g](x_1,\ldots,x_M)=
\prod_{i=1}^M\frac{(\lam^{-2}\zeta^{-2},\lam^2\zeta^{-2},
x_i/\zeta^2 x_{i-1}\phi^2,x_{i-1}\phi^2/x_i\zeta^2;q^2)_\infty}
{(q^2,\zeta^{-4},q^2\lam^2 x_i/x_{i-1}\phi^2,x_{i-1}\phi^2/\lam^2x_i;q^2)_\infty}
\times
}\nonumber\\
&{\ds\times\prod_{i=1}^M \frac{\zeta\sin\pi(2v+t_i-t_{i-1}-h)}{2\pi
i\>\lam x_i}\int_{-i\infty}^{i\infty}ds_1\ldots
\int_{-i\infty}^{i\infty}ds_M\ds\prod_{i=1}^M}
\ds\frac{\ds\biggl(\frac{q^{2+2s_i}\lam
\zeta}{x_{i-1}\phi^2},\frac{q^{2+2s_i}\zeta}{\lam x_i};q^2\biggr)_\infty}
{\ds\biggl(\frac{q^{2s_i}}{\lam
\zeta x_{i-1}\phi^2},\frac{\lam q^{2s_i}}{\zeta x_i};q^2\biggr)_\infty}
\nonumber\\
&\times{\ds\prod_{i=1}^M\frac{\pi\> q^{2s_i}}
{\ds\sin\pi(t_i-s_i+v-I/4)\sin\pi(t_{i-1}-s_i+h-v-I/4)}\>
g(q^{2s_1},\ldots,q^{2s_n})}.\label{pol35}
\end{align}

Repeating the same arguments one can easily show that the integral
operator (\ref{pol35}) has the same polynomial action (\ref{pol27}) as
the $Q$-operator given by (\ref{pol21}). A representation
by the Askey-Roy integral does not require a change of variables
(\ref{pol31}, \ref{pol34}) and probably is more convenient.

We also notice that both  integrals appeared in the study
of orthogonality relations for $q$-Hahn polynomials in \cite{Kaln88}.
It is known that Hahn polynomials appear in the XXX model
as the eigenvalues of the $Q$-operator at $M=2$ \cite{FK95,Kor95,Der99}.
A better understanding of this connection with the approach of \cite{Kaln88}
deserves a further study.

\nsection{$TQ$-relation and commutativity of the $Q$-operators}

In the previous section we constructed the $Q$-operator on the space
of polynomials and proved that it solves the ``left'' $TQ$-relation
\beq
{\bf Q}_f(\lambda){\bf T}_{I}(\lambda;\phi)=
\phi^{ M}\,[\lambda/\zeta]^M {\bf Q}_f(q\lambda)+
\phi^{- M}\,[\lambda \zeta]^M
{\bf Q}_f(q^{-1}\lambda).\label{comm1}
\eeq
on the space of polynomials $K_M[X]$.

The next step is to prove a commutativity of the $Q$-operator with
the transfer-matrix. It immediately follows from a commutativity
of $Q$-operators at different values of the spectral parameters
\beq
[{\bf Q}_f(\lam),{\bf Q}_f(\mu)]=0. \label{comm2}
\eeq

So let us prove the relation 
\beq
{\bf Q}_f(\mu){\bf T}_{I}(\lambda;\phi)={\bf T}_{I}(\lambda;\phi){\bf Q}_f(\mu)
 \label{comm3}
\eeq
as a consequence of (\ref{comm1}-\ref{comm2}). Multiplying both parts of (\ref{comm3}) by 
${\bf Q}_f(\lambda)$ from the left
and using a commutativity (\ref{comm2}) we get
\beq
{\bf Q}_f(\mu){\bf Q}_f(\lam){\bf T}_{I}(\lambda;\phi)=
{\bf Q}_f(\lam){\bf T}_{I}(\lambda;\phi){\bf Q}_f(\mu). \label{comm3a}
\eeq
Substituting the $TQ$-relation (\ref{comm1}) into (\ref{comm3a}) we obtain
\begin{align}
&{\bf Q}_f(\mu)(\phi^{ M}\,[\lambda/\zeta]^M {\bf Q}_f(q\lambda)+
\phi^{- M}\,[\lambda \zeta]^M{\bf Q}_f(q^{-1}\lambda))=\nonumber\\
=&(\phi^{ M}\,[\lambda/\zeta]^M {\bf Q}_f(q\lambda)+
\phi^{- M}\,[\lambda \zeta]^M{\bf Q}_f(q^{-1}\lambda)){\bf Q}_f(\mu) \label{comm3b}
\end{align}
and this again follows from (\ref{comm2}). 

Now let us prove the relation (\ref{comm2}).
As mentioned before
it is sufficient to prove the commutativity of the $Q$-operators
on polynomials. So let calculate
the action of the product of two $Q$-operators on a monomial $x_{i+1}^m$
\beq
{\bf Q}_f(\lam){\bf Q}_f(\mu) \cdot x_{i+1}^m. \label{comm4}
\eeq
Using the action (\ref{pol27}) we can represent the result in the following
form
\beq
{\bf Q}_f(\lam){\bf Q}_f(\mu) \cdot x_{i+1}^m=
\sum_{0\leq n+k\leq m}c(\lam,\mu)_{m,n,k}\> x_{i-1}^n x_{i}^{m-n-k}x_{i+1}^k.
\label{comm5}
\eeq
After straightforward calculations one can derive the following explicit expression
for $c(\lam,\mu)_{m,n,k}$
\begin{align}
&c(\lam,\mu)_{m,n,k}=\phi^{2(m+n-k)}\lam^{2k-m}\mu^{m-2n}
\frac{\ds(\lam^{-2}\zeta^{-2};q^2)_k
(\mu^{-2}\zeta^{-2};q^2)_{m-n}
(\lam^2/\zeta^2;q^2)_{m-n-k}}
{(\zeta^{-4};q^2)_m(\zeta^{-4};q^2)_{m-n}(q^2,\zeta^{-4};q^2)_{n}}\times\nonumber\\
&\times\frac{(q^2;q^2)_m(\lam^2/\zeta^2,\mu^2/\zeta^2;q^2)_n}
{(q^2;q^2)_k(q^2;q^2)_{m-n-k}}
\phifs
\left(\left.\begin{array}{l}\ds
q^{-2(m-k-n)},q^{2+2n-2m}\zeta^4,\frac{1}{\lam^2\zeta^2},
\frac{q^{2n}\mu^2}{\zeta^2}\\
\ds\frac{q^{2n}}{\zeta^4},
\frac{q^{2-2(m-k-n)}\zeta^2}{\lam^2},q^{2+2n-2m}\mu^2\zeta^2
\end{array}\right|q^2,q^2\right),\label{comm6}
\end{align}
where
\begin{align}
\phirs\left(\left.\begin{array}
{l}q^{-n};a_1,\ldots,a_r\\
\phantom{q^{-n},}b_1,\ldots,b_r\end{array}\right|q,z\right)
=\sum_{k=0}^n z^k\,\frac{(q^{-n};q)_k}{(q;q)_k}\prod_{s=1}^r\frac{(a_s;q)_k}
{(b_s;q)_{k}}\,
\label{comm7}
\end{align}
is the standard  terminating basic hypergeometric series.
The series in (\ref{comm6}) also satisfies the balancing condition
$q^{1-n}a_1a_2a_3=b_1b_2b_3$ and $z=q$ (with $q$ replaced by $q^2$).

The commutativity of the $Q$-operators (\ref{comm2})
is equivalent to the symmetry
of coefficients $c(\lam,\mu)_{m,n,k}$
\beq
c(\lam,\mu)_{m,n,k}=c(\mu,\lam)_{m,n,k}. \label{comm8}
\eeq
This symmetry immediately follows from the Sears' transformation (III.15) of
terminating balanced series $\phifs$  in \cite{Gasper}.
Extending (\ref{comm5}) to the action on $x_1^{k_1}\ldots x_M^{k_M}$
we prove (\ref{comm2}) on the space of polynomials $K_M[X]$.

\nsection{Connection between $Q$-operators}

In Sections 2 and 4 we presented two different approaches for construction
of the $Q$-operators in the XXZ spin chain with spin $I/2$ and
arbitrary horizontal
field. The approach of the Section 4 gives much simpler formulas
for the action of one $Q$-operator (\ref{pol27}).
It is easy to see
that this operator maps beyond the finite-dimensional space $V_I$
 and the action (\ref{pol27}) becomes singular  
 for $m>I$, $I\in\mathbb{Z}_+$ due to
a presence of the factor $(q^{-2I};q^2)_m$ in the denominator of
the RHS in (\ref{pol27}).
The $Q$-operators constructed in Section 2
are free from this difficulty.

One can carefully expand near the singular
point $\zeta=q^{I/2}+\epsilon$, $\epsilon\to0$ and extract the invariant
block corresponding to the finite-dimensional subspace $V_I$. This
procedure is  technically challenging and has been completed in \cite{CDKK12a}
for the case of the XXX chain. So we first consider a generic case $I\in\mathbb{C}$.

Let us remind that the $Q$-operators ${\bf A}_+^{(I)}(\lam)$
and ${\bf Q}_f(\lam)$ from Sections 2 and 3
solve the same $TQ$-relation
\beq
{\bf T}_{I}(\lambda;\phi){\bf Q}(\lambda)=
\phi^{ M}\,[\lambda/\zeta]^M {\bf Q}(q\lambda)+
\phi^{- M}\,[\lambda \zeta]^M
{\bf Q}(q^{-1}\lambda).\label{conn1}
\eeq
This $TQ$-relation was obtained from
(\ref{xxz6}) by removing the phase
factor corresponding to different periodicity conditions with respect to
the spectral parameter for two solutions
of (\ref{xxz6}). It implies that two $Q$-operators ${\bf A}_\pm^{(I)}(\lam)$
can not mix (at $\phi\neq1$) and specifying periodicity conditions
 of the solution
we fix it up to a constant matrix multiplier. Therefore,
we must have the relation
\beq
{\bf A}_+^{(I)}(\lam)={\bf A}_0 {\bf Q}_f(\lam), \label{conn2}
\eeq
where ${\bf A}_0$ is some matrix independent of the spectral parameter.

Now let us remind that at $\lam=\zeta$, the $Q$-operator ${\bf Q}_f(\lam)$
 becomes the identity operator
\beq
{\bf Q}_f(\lam)={\bf I}. \label{conn3}
\eeq
It immediately follows from (\ref{conn2}) that
\beq
{\bf A}_0={\bf A}_+^{(I)}(\zeta), \label{conn4}
\eeq
but we already explicitly calculated ${\bf A}_+^{(I)}(\zeta)$
in (\ref{xxz20}). Hence,
we get the following general expression
for the $Q$-operator ${\bf A}_+^{(I)}(\lam)$
\beq
{\bf A}_+^{(I)}(\lam)={\bf A}_+^{(I)}(\zeta){\bf Q}_f(\lam)\label{conn5}
\eeq
at arbitrary values of $\lam$.
Since ${\bf A}_+^{(I)}(\zeta)$ commutes with ${\bf A}_+^{(I)}(\lam)$,
the order of the matrix multiplication in the RHS of (\ref{conn5}) is
irrelevant.

Let us derive the explicit form of the matrix elements of the operator
${\bf Q}_f(\lam)$ in $K_M[X]$. We obtain from (\ref{pol27})
\begin{align}
&{\bf Q}_f(\lam)\cdot x_1^{j_1}\ldots x_M^{j_m}=
\prod_{k=1}^M
\left(\frac{\lam}{\zeta}\right)^{j_k}
\frac{(q^2;q^2)_{j_k}}{(\zeta^{-4};q^2)_{j_k}}\times\nonumber\\
&\times\sum_{l_1=0}^{j_1}\ldots\sum_{l_M=0}^{j_M}
\prod_{k=1}^M
\left(\frac{\phi^2}{\lam^2}\right)^{l_k}
\frac{(\lam^2\zeta^{-2};q^2)_{l_k}(\lam^{-2}\zeta^{-2};q^2)_{j_k-l_k}}
{(q^2;q^2)_{l_k}(q^2;q^2)_{j_k-l_k}}x_k^{j_k-l_k+l_{k+1}}\label{conn6}
\end{align}
Therefore, for the matrix elements of ${\bf Q}_f(\lam)$ we get
\begin{align}
{\bf Q}_f(\lam)_{i_1,\ldots,i_M}^{j_1,\ldots,j_M}&=
\prod_{k=1}^M
\left(\frac{\lam}{\zeta}\right)^{j_k}
\frac{(q^2;q^2)_{j_k}}{(\zeta^{-4};q^2)_{j_k}}\times\nonumber\\
&\times\sum_{l_1=0}^{j_1}
\prod_{k=1}^M
\left(\frac{\phi^2}{\lam^2}\right)^{l_1+\delta_k}
\frac{(\lam^2\zeta^{-2};q^2)_{l_1+\delta_k}
(\lam^{-2}\zeta^{-2};q^2)_{j_k-l_1-\delta_k}}
{(q^2;q^2)_{l_1+\delta_k}(q^2;q^2)_{j_k-l_1-\delta_k}},\label{conn7}
\end{align}
where
\beq
\delta_k=\sum_{s=1}^{k-1}(i_s-j_s). \label{conn8}
\eeq
Comparing (\ref{conn7}) with (\ref{xxz20}) we see that
the divergent factors $(\zeta^{-4};q^2)_{j_k}$ cancel in the matrix product
in the RHS of (\ref{conn5}) and we get the answer for ${\bf A}_+^{(I)}(\lam)$
which is  suitable for a reduction to the finite-dimensional case
$I\in\mathbb{Z}_+$.

Let us emphasize that we derived a general expression for the $Q$-operator
${\bf A}_+^{(I)}(\lam)$ with periodic boundary conditions
based on a highly nontrivial $L$-operator
(\ref{xxz12}) which contain the hypergeometric function $\phit$. To derive
the matrix elements of the $Q$-operator ${\bf A}_+^{(I)}(\zeta)$
we need to multiply $M$ copies of such $L$-operators and to calculate
the trace over the Fock space which seems to be a hopeless problem.

Nevertheless, we now have the explicit expression for ${\bf A}_+^{(I)}(\zeta)$
in terms of the product of two much simpler
matrices ${\bf A}_+^{(I)}(\zeta)$ and ${\bf Q}_f(\lam)$ which are
both finite-dimensional in any subspace $W_l$.
The second $Q$-operator  ${\bf A}_-^{(I)}(\lam)$ for $I\in\mathbb{Z}_+$
can be easily constructed from ${\bf A}_+^{(I)}(\lam)$
using the symmetry (\ref{xxz15}).

It is also interesting to look at the relation (\ref{conn5}) in the limit
$\lam\to\infty$ where ${\bf A}_+^{(I)}(\lam)$ is proportional to the identity 
operator (see (\ref{xxz16})).
In the subspace $W_l$ we obtain
\beq
{\bf Q}_f(\lam)|_{\lam\to\infty}=\lam^l {\bf Q_\infty} (1+O(\lam^{-2})), \label{conn9}
\eeq
where
\beq
[{\bf Q_\infty}]_{i_1,\ldots,i_M}^{j_1,\ldots,j_M}=
\zeta^{-l}\prod_{k=1}^M
\frac{(q^2;q^2)_{j_k}}{(\zeta^{-4};q^2)_{j_k}}
\times\sum_{l_1=0}^{i_1}
\prod_{k=1}^M
\frac{(q^{-2j_k};q^2)_{l_1+\delta_k}}
{(q^2;q^2)_{l_1+\delta_k}}
\left({q^{j_k}\phi }\,{\zeta^{-1}}\right)^{2(l_1+\delta_k)}.\label{conn10}
\eeq
Substituting this into (\ref{conn5}) at $\lam\to\infty$ and using (\ref{xxz16})
we have
\beq
{\bf A}_+^{(I)}(\zeta){\bf Q_\infty}=(-1)^{l+1}\phi^{2M}q^{l-IM}{\bf I},
\label{conn11}
\eeq
i.e. the matrices ${\bf A}_+^{(I)}(\zeta)$ and ${\bf Q_\infty}$
are the inverses of each other up to a simple constant.
In fact, the inversion of the matrix ${\bf Q_\infty}$ in $W_l$
without knowing the explicit answer  (\ref{xxz20})
is  a difficult problem.

Finally, the relations similar to (\ref{conn5}), (\ref{conn11}) were obtained
 in \cite{CDKK12b} for the XXX chain using the operator approach.
So we can look at them as a generalization of results
\cite{CDKK12b} to the case of the XXZ chain.

\nsection{Conclusion}

In this paper we constructed a new integral representation for the XXZ
$Q$-operator at arbitrary complex spin and horizontal field.
In the polynomial basis the action of the $Q$-operator looks surprisingly simple
and the matrix elements are expressed in terms of a single sum.
The generating function of this
action coincides with the generating function for the $Q$-operator  \cite{CDKK13}
constructed as a $q$-difference operator of infinite
order. Our integral $Q$-operator is directly related to the
$Q$-operator ${\bf A}_+(\lam)$ constructed in \cite{Man14} based on the approach
developed in
\cite{BLZ97a,BLZ99a}. Using this connection
 we calculated explicitly
the action of the $Q$-operator ${\bf A}_+(\lam)$ for arbitrary complex spins.
A reduction to (half-) integer spins is non-singular and results in the
explicit formulas for both $Q$-operators ${\bf A}_\pm(\lam)$.

However, when spin is complex a natural duality (\ref{xxz15}) between
${\bf A}_+(\lam)$ and ${\bf A}_-(\lam)$
disappears and construction of the second $Q$-operator
${\bf A}_-(\lam)$ becomes problematic.
It is easy to understand  because
the eigenvalues of ${\bf A}_-(\lam)$ are no longer polynomials. At this
stage we don't know how to construct the
integral analog of ${\bf A}_-(\lam)$ for  complex spins.

Let us notice that in the elliptic case the $Q$-operators for the periodic
XYZ spin chain
constructed in \cite{Zab00} have a natural symmetry between ${\bf Q}_+(\lam)$
and ${\bf Q}_-(\lam)$. However, they were represented as formal infinite series
of difference operators with elliptic coefficients
 and it is difficult to compare their action in any appropriate
basis. It looks like in the trigonometric limit such simple duality disappears and
${\bf A}_-(\lam)$ is a more complicated operator comparing to ${\bf A}_+(\lam)$
in the case of arbitrary spins. This is justified by calculations
in \cite{CDKK12a} for the XXX chain where the action of the second $Q$-operator 
on polynomials has a  more complicated structure.

Another direction of generalization is to extend our results to 
the inhomogeneous case.
Since the kernel of the integral operator ${\bf Q}_f(\lam)$ is factorized, its 
generalization to the inhomogeneous case is straightforward.  
Small lattice calculations (up to $M=3$, $l=3$)
show that the matrix ${\bf A}_0$ in (\ref{conn2}) remains essentially 
the same as in the homogeneous case but no longer commutes with ${\bf Q}_f(\lam)$.
We plan to return to these questions elsewhere.

\section*{Acknowledgments}
I would like to thank Vladimir Bazhanov, Sergey Derkachov,
Michio Jimbo, Anatol Kirillov, Gleb Kotousov, Tetsuji Miwa and Sergey Sergeev
for their interest to this work,
useful discussions and critical comments.
I would also like to thank Jan De Gier for  discussions on
properties of the integral $Q$-operator and collaboration
at the early stages of this project in 2003-2004.
This work is partially supported by the Australian Research Council.

\bibliography{total32m}

\bibliographystyle{utphys}

\end{document}